\documentclass[aps,prl,preprint,groupedaddress,nofootinbib]{revtex4-1}
\usepackage{graphicx}
\begin{document}

%\preprint{}

\title{\large  
 Radiative Seesaw Model with Baryon Number Violation \\ and Upper Limit on Neutron-anti-Neutron Transition Time }


\author{\bf Rabindra N. Mohapatra$^a$}
\author{\bf Nobuchika Okada$^b$}
%\thanks{}
%\affiliation{}
\affiliation{$^a$ Maryland Center for Fundamental Physics and Department of Physics, University of Maryland, College Park, Maryland 20742, USA}
\affiliation{$^b$ Department of Physics, University of Alabama, Tuscaloosa, Alabama 35487, USA}


\date{\today}

\begin{abstract}    
The minimal scotogenic model where small neutrino masses arise  via radiative seesaw, is known to provide a unified framework  for neutrino mass and origin of matter via leptogenesis. However if the inflation reheat temperature of the universe is below the sphaleron reheating temperature, then leptogenesis fails and one way to understand the origin of matter would be to  add an effective interaction involving the right handed neutrino (RHN) $N$ of the form   $\frac{1}{\Lambda^2}N u_Rd_Rd_R$. This model can lead to observable  neutron-anti-neutron ($n-\bar{n}$) oscillation. We show that if RHNs are produced non-thermally,  we can get a cosmological upper limit on the transition time $\tau_{n-\bar{n}}$, which is within the reach of the planned ESS HIBEAM/NNBAR experiment. The proton stability is guaranteed by the scotogenic $Z_2$ invariance, which prevents the appearance of the Dirac mass term for the neutrino.
\end{abstract}

\maketitle

\section{1. Introduction} 

The origin of matter anti-matter asymmetry in the universe is a fundamental mystery of cosmology. Particle physics in the past three decades has opened up several pathways to resolving this mystery, thereby providing a  guide to different new physics scenarios beyond the standard model (BSM). It is therefore important to isolate predictions of these ideas and test them.  We focus on one such direction in this paper.

A popular mechanisms to understand the origin of matter is leptogenesis, which involves the decays of the hypothetical heavy Majorana right handed neutrino (RHN)~\cite{Fukugita:1986hr},  used to understand small neutrino masses 
via the type I seesaw mechanism~\cite{Minkowski:1977sc, Mohapatra:1979ia, GellMann:1980vs, Yanagida:1979as, Glashow:1979nm}.  
The decay process  responsible  is $N\to \ell+\phi$, where $\ell$ and $\phi$ denote, respectively, the left-handed lepton doublet of the standard model and  the scalar doublet responsible for electroweak symmetry breaking. The interesting point about this mechanism is that it uses~\cite{Fukugita:1986hr, Buchmuller:2004nz,  Abada:2006fw, Davidson:2008bu, Bodeker:2020ghk}  the same interactions that  play the key role in understanding the small neutrino masses. 

Within the leptogenesis framework, there are two possibilities: the first possibility is where the RHN is generated by the thermal processes in the early universe. This will happen if the reheat temperature of the universe after inflation is above the RHN masses. 
On the other hand, if the reheat temperature is lower than the RHN mass (i.e. $M_{N_a}> T_R$), the RHNs can only be produced via inflaton decay,  leading to a non-thermal picture for leptogenesis (see for example \cite{Lazarides:1990huy, Murayama:1992ua, Ghoshal:2022kqp, Zhang:2023oyo} ). 

In this paper, we will consider the non-thermal route in a popular neutrino mass model known as the scotogenic model extended to include baryon number (B) violation and study its implications.

The scotogenic model for neutrino mass generation extends the standard model  by adding three right handed neutrinos and an extra scalar doublet which is prevented from having a vacuum expectation value (VEV)~\cite{Tao:1996vb, Ma:2006km, Barbieri:2006dq, Deshpande:1977rw}. The neutrino masses, then  arise as a one loop effect~\cite{Tao:1996vb, Ma:2006km} 
with combined suppression by the loop factor and the RHN masses. This helps to lower the seesaw scale, possibly making it accessible to collider. One can also apply this model to implement thermal leptogenesis using the  $N\to \ell+\eta$ decay as has been discussed in the literature 
(see for instance \cite{Kashiwase:2012xd, Hugle:2018qbw, Borah:2018rca, Sarma:2020msa, Racker:2024fpn, Avila:2025qsc}).

      The cosmology of this model depends however on whether the reheat temperature $T_R$ after inflation is higher or lower than the RHN masses. In the latter case,  in particular, if $T_R \leq T_{sph}\leq M_N$, where $T_{sph}$ is the sphaleron decoupling temperature, neither thermal nor non-thermal  leptogenesis can work to explain the origin of matter. To understand baryogenesis, therefore, we add an effective B-violating interaction of the form  $\frac{1}{\Lambda^2}N u_Rd_Rd_R$, as has already been discussed in the literature~\cite{Dev:2015uca, Davoudiasl:2015jja}. If the RHNs are used in the usual type I seesaw way to get neutrino masses~\cite{Minkowski:1977sc, Mohapatra:1979ia, GellMann:1980vs, Yanagida:1979as, Glashow:1979nm}, then this extended  model leads to rapid proton decay induced by the Dirac mass term in the seesaw matrix and is therefore not phenomenologically viable. On the other hand if one uses the radiative seesaw (or scotogenic seesaw) this problem is avoided~\cite{Dev:2015uca}, since the Dirac mass term connecting $N$ and $\nu$ is forbidden by a $Z_2$ symmetry. Furthermore, the new interaction $\frac{1}{\Lambda^2}N u_Rd_Rd_R$, generates the B-violating process of neutron-anti-neutron oscillation~
 \cite{Kuzmin:1970nx, Glashow:1979nm, Mohapatra:1980qe} which is observable for  10 TeV$-$ 100 TeV scale for $M_N$ and $\Lambda$. This will make the model testable. It is therefore important to discuss the cosmological consequences of this model for baryogenesis.
  
  We find that the model has implications for neutron-anti-neutron oscillation transition time, $\tau_{n-\bar{n}}$, arising from the fact that the lightest member of the second Higgs doublet $\eta$ in the model is unstable.  Making sure that its decay does not affect big bang nucleosynthesis (BBN) and also does not dilute the baryon asymmetry, leads to upper limits on  $\tau_{n-\bar{n}}$, keeping it in accessible range for  the upcoming HIBEAM/NNBAR experiment planned at ESS.
To the best of our knowledge, this is the second model where $n-\bar{n}$ oscillation is directly responsible for the origin of matter, with the model leading to an upper bound on the $n-\bar{n}$ oscillation time (the other model \cite{Babu:2006xc} being the post-sphaleron baryogenesis  model  with $S\to u_Rd_Rd_Ru_Rd_Rd_R$ decay~\cite{Babu:2013yca}, where $S$ is a real scalar field.). We then provide a high scale UV-completion of the model and show how proton decay is forbidden by the same $Z_2$ symmetry that leads to scotogenic neutrino mass generation. 
 
 The paper is organized as follows: in sec.~2, we present the effective Lagrangian of the model with baryon number violating RHN coupling; in sec. 3, we discuss the cosmology of the model, addressing questions such as baryogenesis; in sec. 4, we discuss cosmological limits on  $\tau_{n-\bar{n}}$. In sec. 5, we discuss constraints on the model from baryon number washout. In sec. 6, we 
 present a renormalizable UV complete version of the model. In sec. 7 we add some comments on the model, and we conclude in sec. 8.

 %in sec.~3, we discuss how the non-thermal origin of RHNs in the early universe prevents successful leptogenesis and find the constraint of resonant baryogenesis on the model properties. In sec.~4, we discuss how $n-\bar{n}$  oscillation arises as a signature of the model with an upper limit on its oscillation time. Sec.~5 is devoted to a discussion of constraints from baryon asymmetry and dark matter survival and that of dark matter detection constraints. In sec.~6, we  We close with some comments in sec.~7 and conclusions in sec.~8.

\section{\bf 2. The scotogenic  Lagrangian with $B$-violation}  
The scotogenic model  for neutrino mass~\cite{Tao:1996vb, Ma:2006km} is an extension of the standard model, which adds to it three heavy Majorana neutrinos $N_i$ ($i=1,2,3$) (called RHN below) and a new $SU(2)_L$ doublet, $\eta$ having no VEV and interacting only with the singlet neutrinos $N$ as $Y_D\bar{ \ell}\eta N$. This results from a new $Z_2$ symmetry which allows only this new interaction and Majorana mass terms for the $N$  as well as the extended two Higgs potential discussed below. Since the $\eta$ doublet does not have a VEV, the model does not have a Dirac neutrino mass term and hence, does not give the usual tree-level seesaw.  Neutrino mass arises at the one loop level. 
The original model has no B-violation. %Therefore, the only way to understand the origin of matter is via leptogenesis.

The $Z_2$ invariant Higgs potential of the model is given by 
\begin{eqnarray}
V(\phi, \eta)~=~&&-\mu^2 \phi^\dagger \phi +m^2 \eta^\dagger \eta+ \frac{\lambda_1}{2} (\phi^\dagger \phi )^2
+\lambda_2 ( \eta^\dagger \eta)^2\\\nonumber
&&+\lambda_3 (\phi^\dagger \phi)  (\eta^\dagger \eta) +\lambda_4(\phi^\dagger \eta)(\eta^\dagger \phi)+\frac{1}{2}\lambda_5[(\phi^\dagger \eta)(\phi^\dagger \eta)+h.c.], 
\end{eqnarray}
where $\phi$ is the SM Higgs doublet. 
To this theory, we add a new gauge singlet complex field $\Phi$ to play the role of inflaton which couples to two RHNs and
 a new coupling of $N$ to three right handed quarks so that the relevant effective Lagrangian for the model in the RHN sector looks as follows: 
\begin{eqnarray}
{\cal L}_I~=~Y_D^{i j} \overline{\ell_i}\eta N_j- M_{i} \, \overline{N_i^{{}C}}N_i + \frac{\kappa_i}{\Lambda^2}N_i u_Rd_Rd_R
+\lambda_0\Phi \overline{N_i^{{}C}}N_i +h.c. ,
\label{eqL}
\end{eqnarray}
where $\ell$ is the left-handed lepton doublet of SM, and $\kappa_i$ are complex phases.
We will spell out the detailed action of the $Z_2$ symmetry that makes the above theory natural when we present the UV-complete Lagrangian in a subsequent section.

Successful inflation can be implemented via the following non-minimal coupling of $\Phi$ to gravity:
\begin{eqnarray}
{\cal L}_g~=~-\frac{1}{2}(M^2_P+2 \, \xi\,  \Phi^\dagger \Phi)R.  %-\lambda_{\Phi}( \Phi^\dagger \Phi)^2.
\end{eqnarray}
A suitable choice of the coupling constant $\xi >0$ leads to a successful inflation scenario (see, for example, \cite{Okada:2010jf}). 

Using the fact that the only field that has a VEV is the SM Higgs doublet $\phi$, the masses of the physical scalars are given by~\cite{Deshpande:1977rw}:
\begin{eqnarray}
 &m^2_h&~=~ \lambda_1 v^2 ,\nonumber\\
 &m^2_{\eta^+}&~=~m^2+ \frac{1}{2}\lambda_3 v^2 , \nonumber\\
 &m^2_{\eta^0_R}& ~=~m^2+ \frac{1}{2}(\lambda_3+\lambda_4 +\lambda_5)v^2 , \nonumber\\
 &m^2_{\eta^0_I}& ~=~m^2+ \frac{1}{2}(\lambda_3+\lambda_4 -\lambda_5)v^2 , 
\end{eqnarray}
where $\langle \phi^0 \rangle =\frac{v}{\sqrt{2}}$, and $\eta^0_{R,I}$ are real and imaginary parts of the $\eta^0$ field.
In what follows, we will denote $\eta_R\equiv \eta$ and choose $\lambda_5 < 0$ so that it is the lightest $Z_2$-odd particle in the model, as noted above . 

\subsection{2.1 Phenomenology of the new  physical scalars in the model}

The implications of the $\eta$-fields for the Large Hadron Collider searches have been discussed and various constraints have been derived on $\lambda_{3,4,5}$ ~\cite{Lundstrom:2008ai, Ilnicka:2015jba, Lahiri:2025opz}, which parameterize their masses and couplings. They depend sensitively on the relative mass differences among the component fields in the $\eta$ doublet.
The charged scalar $\eta^\pm$ can be produced at LEP via the Drell--Yan process,
$e^+ e^- \to \eta^+ \eta^-$, mediated by the photon and the $Z$ boson. 
From the LEP~2 results, one obtains the lower bound $m_{\eta^\pm} > 70~\text{GeV}$ \cite{Pierce:2007ut}. 
The charged scalar $\eta^\pm$ can also be produced at the LHC. 
Once produced, it decays as $\eta^\pm \to W^\pm + \eta$. 
This production and decay topology are analogous to chargino ($\tilde{\chi}^\pm$) pair production followed by
$\tilde{\chi}^\pm \to W^\pm + \tilde{\chi}_1^0$ in the minimal supersymmetric standard model, where
$\tilde{\chi}_1^0$ is the lightest neutralino. 
Although the production cross section for an $\eta^\pm$ pair is smaller than that for chargino pair production by a factor of a few,
it is expected that the LHC sparticle search results impose similar constraints on the inert doublet model.
For example, the ATLAS Collaboration has reported sparticle search results using up to $140~\text{fb}^{-1}$ of integrated luminosity
at $\sqrt{s}=13~\text{TeV}$ \cite{ATLAS:2024lda}. 
If the mass difference between the chargino and the neutralino is large, the lower bound on the chargino mass is found to be around
$800~\text{GeV}$.
At the LHC, associated production of $\eta_I \, \eta^\pm$ can also occur through a $W$-boson mediated process,
followed by the decays $\eta_I \to \eta + Z$ and $\eta^\pm \to W^\pm + \eta$.
This process is analogous to $\tilde{\chi}_2^0 \, \tilde{\chi}^\pm$ production, followed by
$\tilde{\chi}_2^0 \to \tilde{\chi}_1^0 + Z$ and $\tilde{\chi}^\pm \to W^\pm + \tilde{\chi}_1^0$,
where $\tilde{\chi}_2^0$ is the second-lightest neutralino.
Searches for three leptons plus missing transverse energy arising from leptonic decays of the $W$ and $Z$ bosons
yield even stronger constraints: assuming the lightest neutralino is much lighter,
the lower bound on the chargino mass reaches approximately $1~\text{TeV}$ \cite{ATLAS:2024lda}.
These severe constraints, however, do not apply when the lightest neutralino is nearly degenerate with the chargino.
For example, for $m_{\tilde{\chi}^\pm} - m_{\tilde{\chi}_1^0} \simeq 1~\text{GeV}$, only the LEP bound is relevant \cite{ATLAS:2024lda}.

%In our analysis, we therefore choose model parameters such that
%$m_{\eta^\pm} - m_\eta \simeq 1~\text{GeV}$ and $m_{\eta^\pm} > 70~\text{GeV}$.
%As discussed below, we impose the condition $\lambda_3 + \lambda_4 + \lambda_5 = 0$,
%which switches off the coupling between the dark matter candidate $\eta$ and the Standard Model Higgs boson and keeps the model compatible with the latest DM search results (see for instance~\cite{LZ:2024zvo}).
%As a result, the small mass splitting  between $m_{\eta^\pm} - m_\eta \simeq 1~\text{GeV}$ is controlled by a small value of $\lambda_3$.
%In Fig.~\ref{fig:MassSpectrum}, we show the inert doublet mass spectrum as a function of $m_\eta$.
%For illustration, we take $\lambda_3 = 0.006$, $\lambda_4 = 0.294$, and $\lambda_5 = -0.3$,
%which satisfy $\lambda_3 + \lambda_4 + \lambda_5 = 0$.
Note  that when perturbativity of the quartic couplings is imposed up to the inflationary or Planck scale,
their magnitudes are typically bounded by $|\lambda_i| \lesssim 0.3$.
See, for example, Ref.~\cite{Chowdhury:2015yja} for a detailed renormalization group analysis.
For our choice of parameters, we have explicitly checked that all running couplings $\lambda_i$
remain in the perturbative regime.
Moreover, the running quartic coupling $\lambda_1$ of the Standard Model Higgs doublet remains positive for $|\lambda_{3,4,5}|\gtrsim 0.1$,
ensuring the stability of the electroweak vacuum.

Furthermore, we demand that the masses of the $\eta$ fields be less than a TeV to avoid destabilizing the SM Higgs boson mass,  since the one loop contribution to the Higgs boson mass in the theory is given by
\begin{eqnarray}
\Delta m^2_\phi\sim \frac{\lambda_{3,4}}{16\pi^2} m^2_\eta.
\end{eqnarray}
This is the naturalness bound on $\eta$-masses.

To close this section, we note that the lightest $\eta$ particle is unstable unlike in the original scotogenic model and decays to $\eta\to udd\nu$ with a decay width
\begin{eqnarray}
\Gamma_\eta \simeq 1.3\times 10^{-7}\frac{Y^2_D m^7_\eta}{M^6_N\beta^4} ,
\end{eqnarray}
where $\beta= \Lambda / m_\eta$, and clearly within our assumption of effective B-violating Lagrangian $\beta \gg 1$. The $\eta$ field could be fairly long lived with lifetime close to one second or less. This will have consequences for the $n-\bar{n}$ transition, as we discuss later. 

%%%%%%%%%%%%%

\section{3. Cosmology of the model}

The cosmology of the model goes as follows:  for values of $\Phi$ of order or higher than the Planck mass, $M_P$, the potential $V(\Phi)$ in the Einstein frame is almost a constant as a function of $\Phi$, giving rise to inflation. As $\Phi$ rolls down, the inflaton potential  $V(\Phi)$  is no more a constant and inflation ends~\cite{Bezrukov:2007ep}. After this the inflation field decays causing reheating of the universe. 
We arrange parameters such that the inflaton decay width $\Gamma_{\Phi\to NN} \ll \Gamma_N$ so that once $N$ is created from the decay of inflaton, it immediately decays and reheats the universe. Thus production of $N$ is a non-thermal process.  The number density of $N$ subsequent to $\Phi$ decay, $n_N$, is then equal to twice $n_\Phi$, the number density of the inflaton.

\subsection{3.1 Failure of non-thermal Leptogenesis}

Baryogenesis in this model can take place either via $N\to \ell +\eta$ decay in combination with sphalerons or via $N\to 3q$ decay.  If however, the reheat temperature $T_R$  after inflation is below the sphaleron decoupling temperature $T_{sph} \simeq 130$ GeV, 
which we assume here, leptogenesis does not work and matter-anti-matter asymmetry can only arise from the  $N \to 3q$ decay in combination with CP violation.

%, when we have:
%\begin{eqnarray}
%n_L\simeq 2 \, n_\Phi \, Br (N\to \ell+\eta) \, \epsilon_L, 
%\end{eqnarray}
%where $\epsilon_L$ is the lepton asymmetry parameter generated by the $N$-decay. 
%We estimate $n_\Phi\simeq \frac{\rho_\Phi}{m_\Phi}$, where $\rho_\Phi$ is the inflaton energy density prior to its decay. Using the expression for entropy at $T=T_R$, we deduce that
%\begin{eqnarray}
%\frac{n_L}{s}= \frac{3T_R}{2m_\Phi}Br (N\to \ell+\eta) \epsilon_L
%\end{eqnarray}
%
%We will now put the constraint of adequate dark matter generation (Eq.~(\ref{DMmass})) on this model. 
%If leptogenesis is the mechanism for baryogenesis, then  $Y_\eta \epsilon_L\simeq 10^{-10}$ which implies

\subsection{3.2 Baryogenesis via $N\to u_Rd_Rd_R$ decay} 

The baryon number generated by $N$ decay is given by 
\begin{eqnarray} 
  n_B = n_N \, Br(N \to 3 q) \, \epsilon_B  %=2 n_\Phi \, Br(N \to 3 q) \, \epsilon_B 
   =2 \frac{\rho_\Phi}{m_\Phi }\, Br(N \to 3 q) \, \epsilon_B ,
\end{eqnarray} 
where $\epsilon_B$ is the baryon asymmetry parameter that arises from tree and one loop interference of diagrams for $N\to 3q$ decay, 
  and we have used $n_N=2 n_\Phi =  \frac{\rho_\Phi}{m_\Phi }$ with the inflaton mass $m_\Phi$ and 
  $\rho_\Phi$ being the inflaton energy density. 
Assuming $\eta$ to be out of equilibrium, the radiation energy density created by $N$ decay at the time of inflaton decay 
  is estimated as 
\begin{eqnarray}
        \rho_{rad} = \rho_\Phi \left( 1-  \frac{1}{2}Br(N \to \eta \ell)\right) =\frac{\pi^2}{30} \, g_* \,T_R^4, 
\end{eqnarray}
where $g_* \simeq 100$ is the effective degrees of freedom of relativistic particles in thermal plasma. 
Using $\rho_{rad}/s = 3/4 T_R$ for the thermal plasma, we estimate the baryon asymmetry generated by the $N \to 3q$ decay as 
\begin{eqnarray}
  \frac{n_B}{s}\simeq \frac{3T_R}{2m_\Phi} \, \frac{ Br(N\to 3q)}{1-\frac{1}{2}Br(N\to \eta\ell)} \, \epsilon_B .
 \label{YB}  
\end{eqnarray}

Similarly, we can derive an expression for $Y_\eta\equiv  \frac{n_\eta}{s}$ :
\begin{eqnarray}
Y_\eta(t=\tau_\Phi)~=~\frac{3T_R}{2m_\Phi} \, \frac{Br(N\to \eta \ell) }{1-\frac{1}{2} Br(N\to \eta\ell)}. 
 \label{Y_eta}
\end{eqnarray}
Combining  Eqs.~(\ref{YB}) and (\ref{Y_eta}), we can express $Y_\eta$ in terms of the parameters of the theory:
\begin{eqnarray}
Y_\eta   \simeq  \frac{10^{-10}}{\epsilon_B} \,  \frac{Br(N\to \ell+\eta)}{Br(N\to 3q)}, 
\end{eqnarray}
where we have used the observed value, $n_B/s \simeq 10^{-10}$.  
In the above equation, the asymmetry parameter $\epsilon_B$  is given by \cite{Dev:2015uca, Davoudiasl:2015jja}
 \begin{eqnarray}
 \epsilon_B\simeq \frac{Im( \kappa^{2*}_1\kappa^{2}_2)}{3072\pi^3|\kappa_1|^2}\left(\frac{M_1}{\Lambda}\right)^4\frac{M_1M_2}{M^2_2-M^2_1}.
 \end{eqnarray}
 where we have denoted RHN masses for different generations  $M_{1,2,3}$ with at least two of them being quasi-degenerate to facilitate an adequate value for $\epsilon_B$.

The partial decay widths of $N \to \ell+\eta$ and $N \to 3 q$ are, respectively, estimated as 
\begin{eqnarray}
 \Gamma_{N\to  \ell+\eta} \simeq  \frac{Y^2_D}{8 \pi}M_N , \; \; \;
 \Gamma_{N\to 3q} \simeq \frac{M_N^5}{192 \,\pi^3 \,\Lambda^4} .
\end{eqnarray}
Using $\frac{Br(N\to \eta \ell)}{Br(N\to 3q)}= \frac{\Gamma_{N\to  \ell+\eta} }{ \Gamma_{N\to 3q} }$, 
% \simeq 24\pi^2 \, Y^2_D \, \beta^4 $, 
we get
  \begin{eqnarray}
  Y_\eta  \simeq  \frac{10^{-10}}{\epsilon_B} 24\pi^2 \, Y^2_D \, \beta^4 .
\label{Yeta}  
  \end{eqnarray}
  This gives us an expression for the energy density of $\eta$ field when it decays ($t=\tau_\eta$) as
  \begin{eqnarray}
  \rho_\eta(t=\tau_\eta)= m_\eta \, Y_\eta \, s(t=\tau_\eta) , 
  \end{eqnarray}
where $\tau_\eta$ is the lifetime of $\eta$, and $s(t=\tau_\eta)$  is the entropy density of the universe at $t=\tau_\eta$.

\section{4. Theoretical limits on $\tau_{n-\bar{n}}$ in the model}

To convert the above results  into a bound on $n-\bar{n}$ transition time,  we note that $n-\bar{n}$ oscillation in the model arises from the exchange of Majorana $N$ between two $u_Rd_Rd_R$ vertices. The resulting $\Delta B=2$ interaction has the strength:
\begin{eqnarray}
{\cal L}_{\Delta B=2}~=~ \frac{1}{\Lambda^4 M_N}u_Rd_Rd_Ru_Rd_Rd_R +h.c.
\end{eqnarray}
Clearly, this is directly connected to the Majorana mass of the neutrinos.
After hadron dressing, we find the $\tau_{n-\bar{n}}$ transition time to be
\begin{eqnarray}
\tau_{n-\bar{n}}[s] \simeq \frac{M_N \, \Lambda^4}{|{\cal M}|}\times 6.6\times 10^{-25} \simeq 10^{-20} M_N^5 \beta^4, 
\label{tauNN}
\end{eqnarray}
where $M_N$ is given in units of GeV, and ${\cal M}=-6.5\times 10^{-5}$ GeV$^6$ from a lattice QCD calculation~\cite{Rinaldi:2019thf}. 
Similar  hadron dressing factors  $|{\cal M}|$ also arise in early  bag model calculations~\cite{Rao:1982gt}.
Proton decay is forbidden in the model since $\eta$-field does not have a VEV.

\subsection{4.1 Constraints from the decay time of $\eta$}
\begin{figure}[t!]
        \centering
	    \includegraphics[scale=0.6]{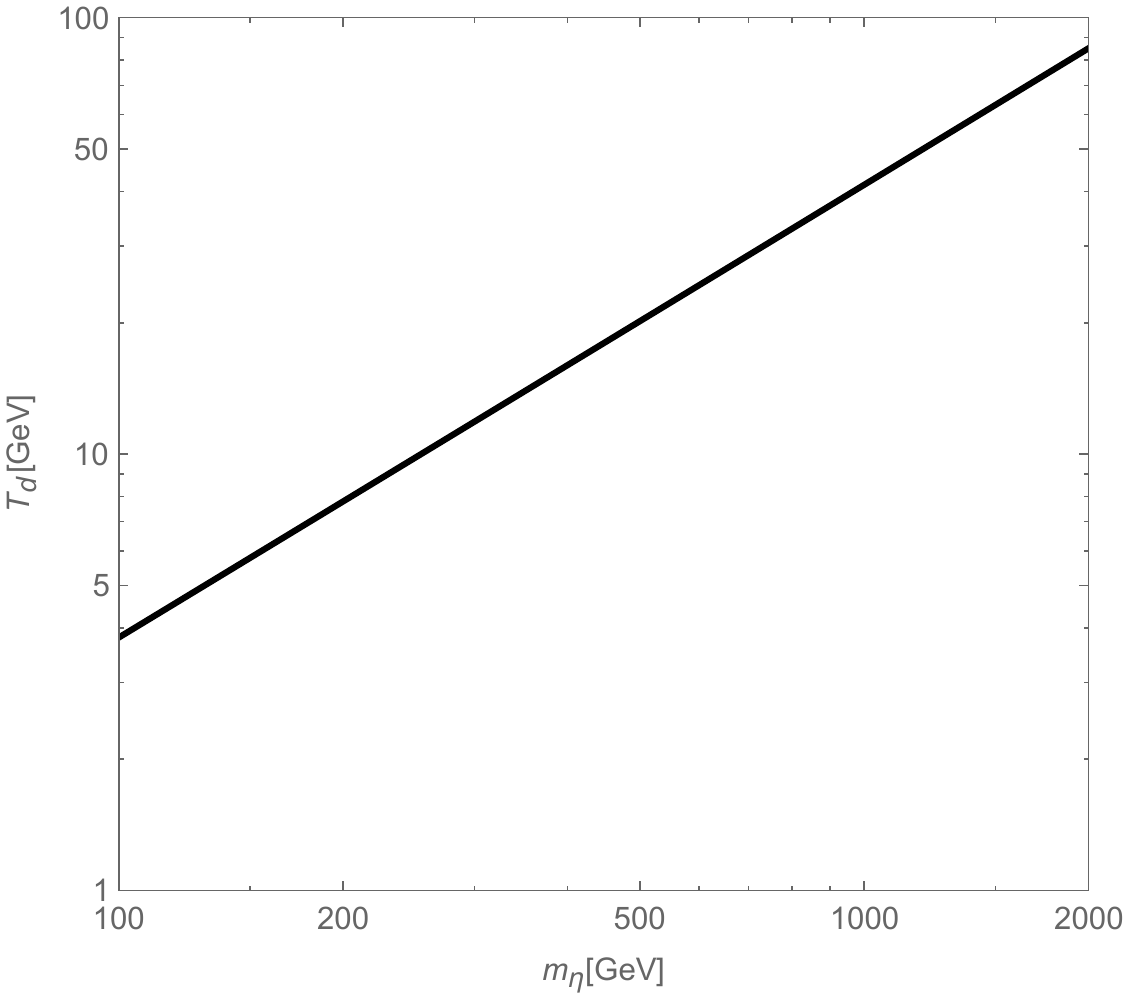}
	%\end{subfigure}
 \caption{ Profile of the $T_d$ as a function of $\eta$ mass. We find that $m_\eta/T_d\simeq 20-27$. 
 %As a comparison, we also show the sphaleron decoupling temperature of about $T_{sph}=130$ GeV (horizontal dashed line). 
 }
\label{fig:Td}
\end{figure}

%%%%%%%%%%%
Note that the new neutral scalar field $\eta$ has  interactions mediated by $W, Z$  bosons which keep it in equilibrium until quite low temperature. The processes of concern here are $\eta+\eta\to WW, ZZ, hh$ which deplete its density 
whereas the inverse processes get it into thermal equilibrium. 
The temperature at which $\eta$ fully decouples from the cosmic soup can be determined  by demanding that
\begin{eqnarray}
n_\eta \langle \sigma (\eta\eta\to SM)v_{rel} \rangle \simeq  \frac{n_\eta(T_d) }{4\pi m^2_\eta}
=\sqrt{\frac{\pi^2g_*}{90}}\frac{T^2_d}{M_P}. 
\end{eqnarray}
Taking $n_\eta (T_d)\simeq\left(\frac{m_\eta T_d}{2\pi}\right)^{3/2} e^{-m_\eta/T_d}$ , $g_*\simeq 100$ 
and $M_P\simeq 2.4\times 10^{18}$ GeV, we find $T_d$ and plot it as a function of $m_\eta$ in Fig.~\ref{fig:Td}. 
For $\eta$ to be non-thermal, we must have $T_d >  T_R$.  The naturalness bound of $m_\eta < $ TeV then implies  from
Fig.~\ref{fig:Td} that the reheat temperature of the universe in our model must be at or  below 10 GeV. 

%%%%%%%%%%%%%%%%%%%
%%%%%%%%%%%%%%%

Due to its large mass, $\eta$ cloud carries a lot of energy which is released when it decays. This affects the big bang cosmology. 
If $\eta$ decays after BBN, the energetic decay products may destroy light nuclei successfully synthesized in the BBN epoch. 
%it will increase the expansion rate of the universe at the BBN epoch and will change the ratio of neutrons to protons 
%at that time which in turn will change the helium to hydrogen ratio in disagreement with observations. 
We therefore demand that $\tau_\eta \gtrsim 1$ sec, or equivalently, the temperature of the thermal plasma $T_{decay}$ 
when $\eta$ decays should be above a few MeV. 

For $m_\eta < M_N$, $\eta$ decays to $\nu +3 q$ mediated by the Majorana RHN, and this decay width is estimated as 
\begin{eqnarray}
  \Gamma_\eta \simeq PS^{(4)} \frac{Y_D^2 \, m_\eta^7}{M_N^2 \, \Lambda^4} 
   \simeq 1.3\times 10^{-7} \times \frac{Y_D^2 \,m_\eta^7}{M_N^6 \, \beta^4},  
\end{eqnarray}
where we estimate the four body phase space factor $PS^{(4)}=\frac{1}{2(4\pi)^5 \Gamma(4)\Gamma(3)}\simeq 1.3\times 10^{-7}$.  
Using $\Gamma_\eta =H(T_{decay})$, we find 
\begin{eqnarray}
T_{decay}=\left(1.3\times 10^{-7} \times \sqrt{\frac{90}{\pi^2 g_*}} \, 
\frac{Y^2_D \, m^7_\eta}{M^6_N \, \beta^4} \, M_P \right)^{1/2} .
\label{Tdecay}
\end{eqnarray}
We use an approximate expression for one-loop $m_\nu$ in this model,
 \begin{eqnarray}
 m_\nu\simeq \frac{|\lambda_5| \, Y^2_D \, v^2}{16 \, \pi^2 \, M_N},
 \end{eqnarray}
with $m_\nu \simeq 5 \times 10^{-11}$ GeV for a typical neutrino mass scale from the neutrino oscillation data 
to replace $Y^2_D=2.6\times 10^{-13}\frac{M_N}{|\lambda_5|}$. 
Here, $M_N$ is given in units of GeV. 
Using the fact that $\tau_{n-\bar{n}}\simeq 10^{-20} \, ({\rm sec.})M^5_N\beta^4$ in this model, we get
\begin{eqnarray}
\tau_{n-\bar{n}}[{\rm s}]=\ \simeq 2.4\times 10^{-22} \, \frac{m^7_\eta}{|\lambda_5| \, T^2_{decay}}, 
\end{eqnarray}
where $m_\eta$ and $T_{decay}$ are given in units of GeV. 

\begin{figure}[t!]
        \centering
	    \includegraphics[scale=0.5]{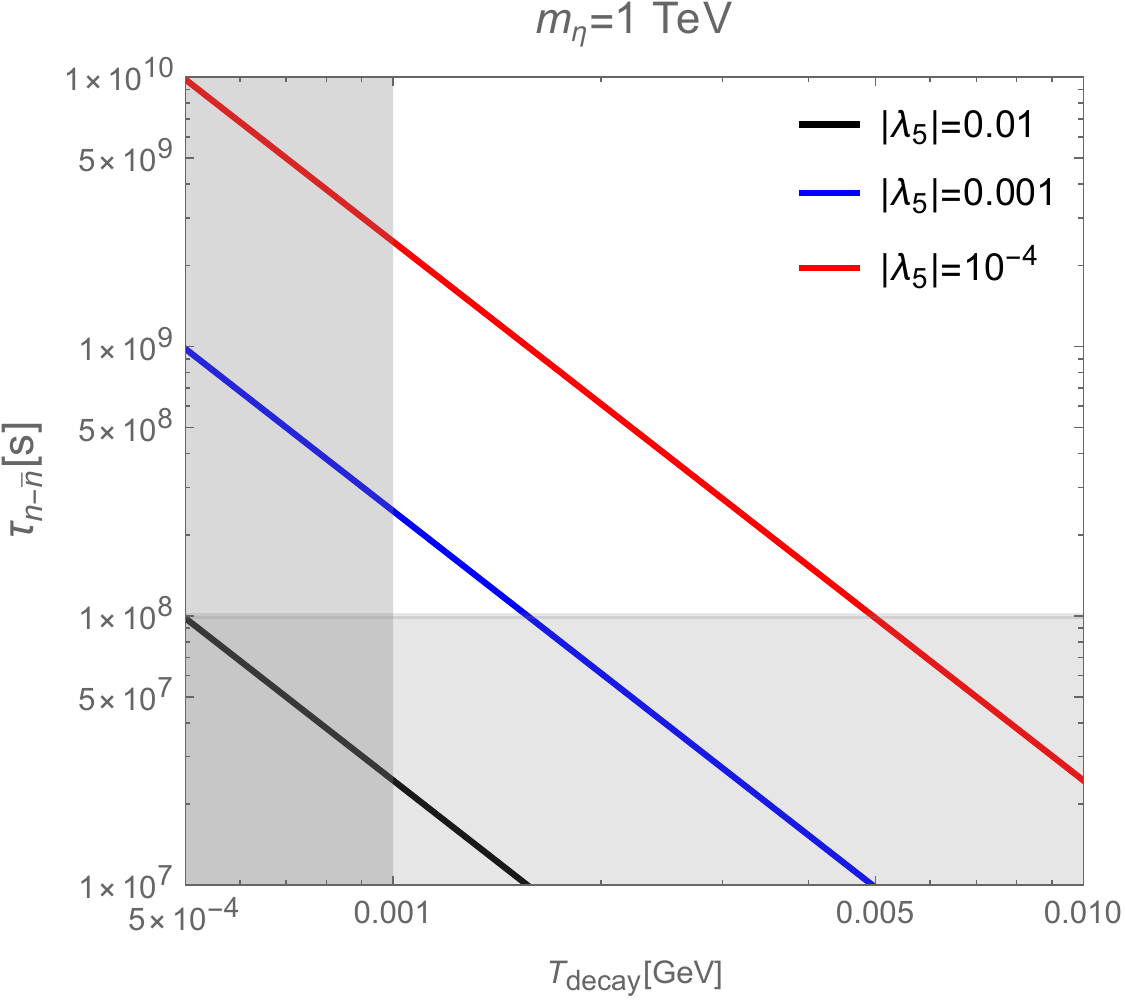}
	%\end{subfigure}
 \caption{
 The predicted free neutron-anti-neutron oscillation time $\tau_{n-\bar{n}}$ is shown as a function of $T_{decay}$ 
 for several values of $|\lambda_5|$. The gray shaded region is excluded by the current experimental limit, 
 $\tau_{n-\bar{n}} > 10^{8}\ \mathrm{sec.}$ ~\cite{Baldo-Ceolin:1994hzw, Super-Kamiokande:2020bov} and $T_{decay} > 1$ MeV from BBN.
}
\label{fig:TauNN-1} 
\end{figure}

 The limits on $\tau_{n-\bar{n}}$ for plausible parameter ranges of the model are shown in Fig.~\ref{fig:TauNN-1}.  We see that most of the parameter range of the model, the predicted upper limits are within the accessible range of the planned HIBEAM/NNBAR experiment at ESS. We also note from Fig.~\ref{fig:TauNN-1}  that higher values of $|\lambda_5|$ lead to lower $\tau_{n-\bar{n}}$ , that are in conflict with current experimental limits. Thus,$\eta$  lifetime considerations  prefer lower values of  $|\lambda_5|$.
  %%%%%%%%%%%%%%%%%%
%%%%%%%%%%%%%%%%%%
 
 \subsection{4.2 Limits from dilution effects on $n_B$ from the decay of $\eta$}
 
Since $\eta$ is a long-lived particle, it may dominate the energy density of the universe when it decays. 
If so, the baryon asymmetry created before the decay is diluted. 
To avoid the dilution, we must therefore demand that $\rho_\eta (t=\tau_\eta) < \rho_{rad}(t=\tau_\eta) $. 
This condition translates to $m_\eta Y_\eta s(T_{decay} ) < \rho_{rad}(T_{decay})$, or equivalently, 
 \begin{eqnarray}
 m_\eta Y_\eta < \frac{3}{4} T_{decay} .
 \end{eqnarray}
 Combining this with $Y_\eta$ from Eq.~(\ref{Yeta}), $T_{decay}$ from Eq.~(\ref{Tdecay}) and 
 $Y^2_D=2.6\times 10^{-13}\frac{M_N}{|\lambda_5|}$, the condition translates to
 \begin{eqnarray}
 \frac{m_\eta Y_\eta}{3/4 T_{decay}}~=~5.2\times 10^{-20}\frac{\beta^6}{\epsilon_B}\sqrt{\frac{M^7_N}{m^5_\eta|\lambda_5|}}< 1.
 \end{eqnarray}
 %Defining $M_\eta/M_N`=~r$ (so that $r<1$) and using $\tau_{n-\bar{n}}=10^{-20} M^5_N\beta^4$, Eq. (18) translates to
% \begin{eqnarray}
 %  \tau_{n-\bar{n}} <\frac{2.5\times 10^{76}}{\beta^{26}}\epsilon^5_B r^{25/2} |\lambda_5|^{5/2}
  % \end{eqnarray}
Applying this condition to Eq.~(\ref{tauNN}), we get an upper bound on $\tau_{n-\bar{n}}$: 
 \begin{eqnarray}
 \tau_{n-\bar{n}}[{\rm s}] ~<~\frac{3.5\times 10^7 \, m^{25/7}_\eta \, \epsilon^{10/7}_B \, |\lambda_5|^{5/7}}{\beta^{32/7}}, 
 \end{eqnarray}
where $m_\eta$ is given in units of GeV. 
 
\begin{figure}[t!]
        \centering
	    \includegraphics[scale=0.5]{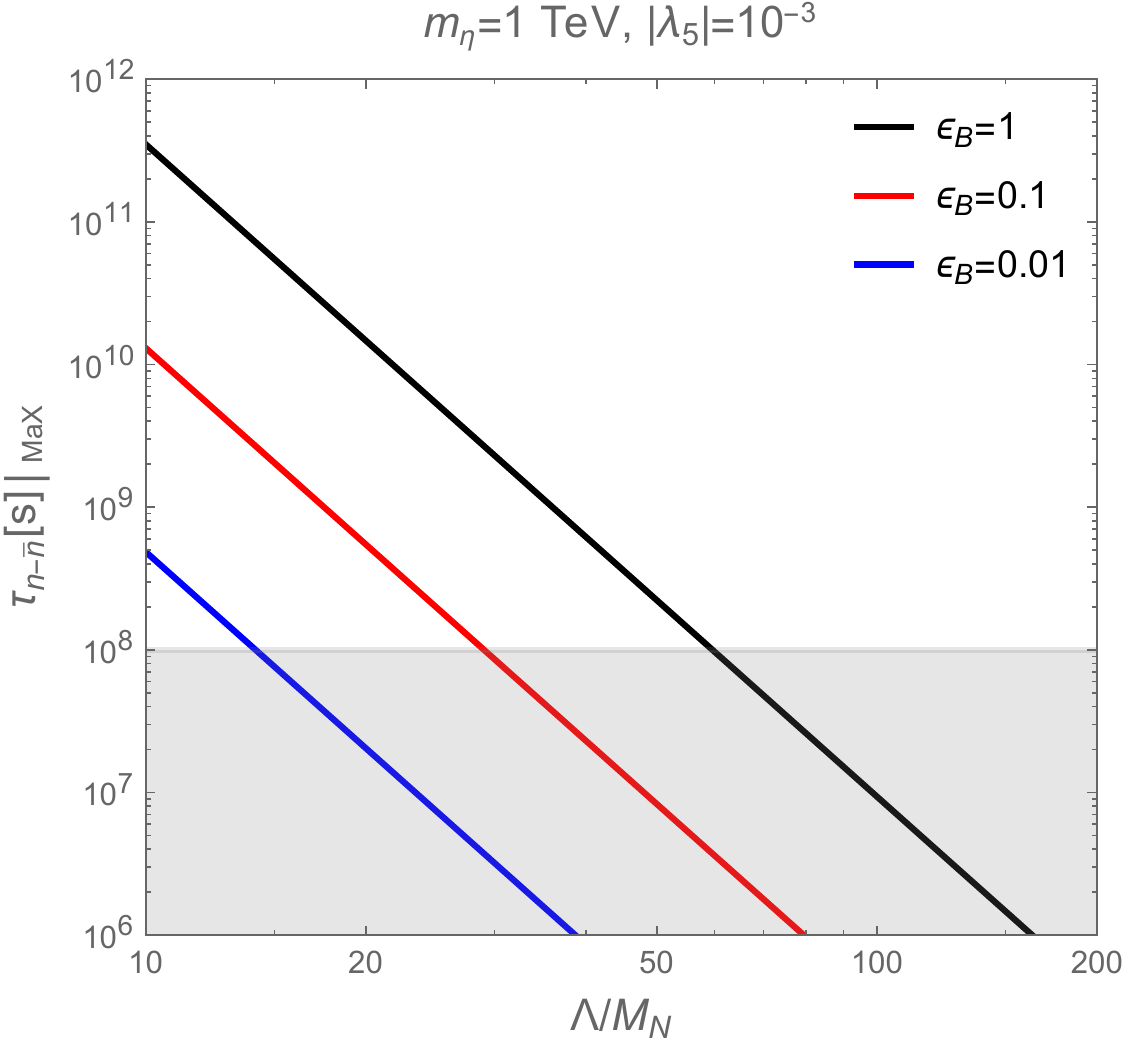}
	%\end{subfigure}
 \caption{Theoretical upper limit on $\tau_{n-\bar{n}}$ as a function of $\frac{\Lambda}{M_N}$ for $| \lambda_5| =10^{-3}$ and various values of $\epsilon_B$.}
\label{fig:TauUB-1}
\end{figure}

\begin{figure}[t!]
        \centering
	    \includegraphics[scale=0.5]{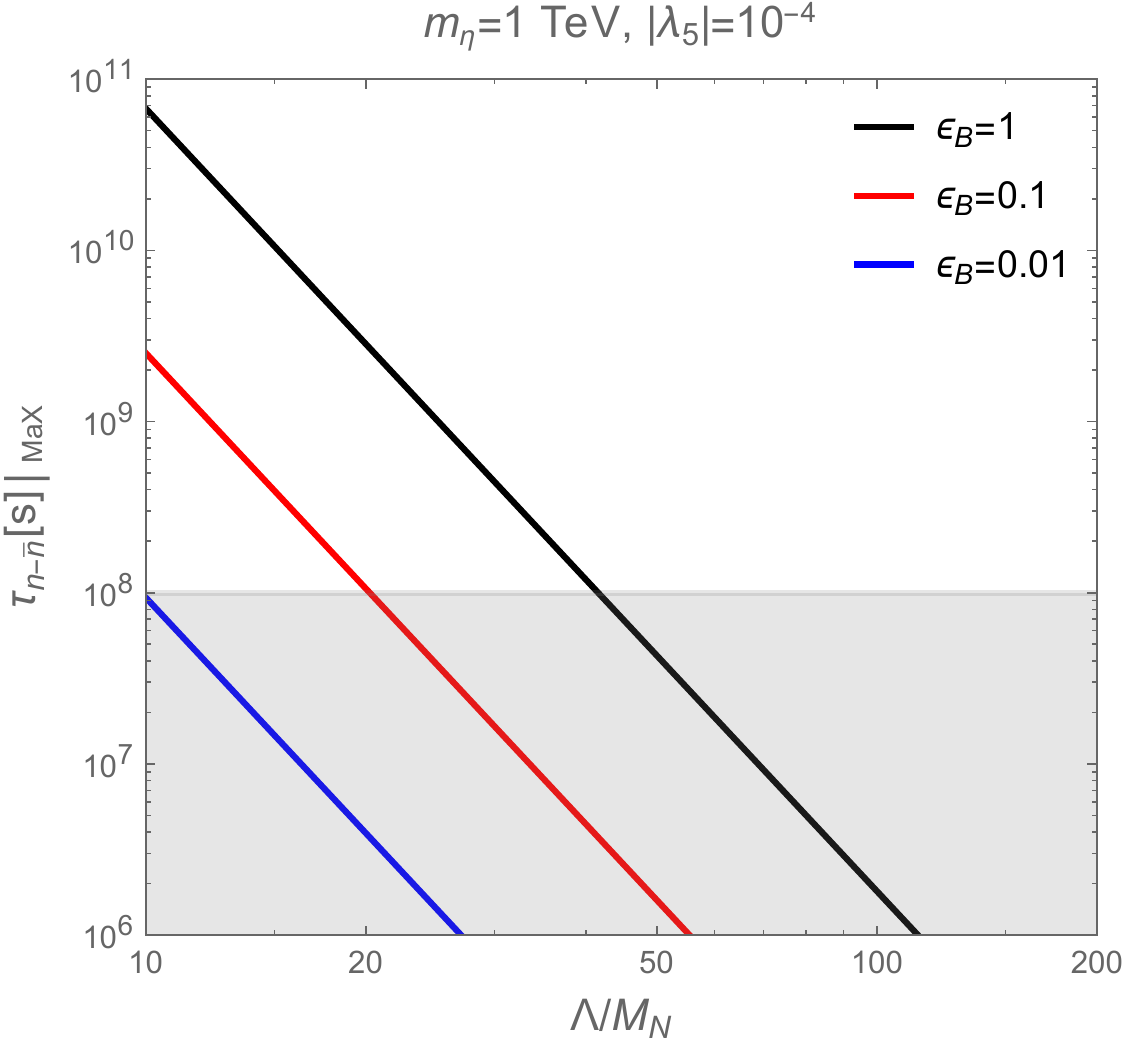}
	%\end{subfigure}
 \caption{Theoretical upper limit on $\tau_{n-\bar{n}}$ as a function of $\frac{\Lambda}{M_N}$ for $| \lambda_5| =10^{-4}$ and various values of $\epsilon_B$ }
\label{fig:TauUB-2}
\end{figure}

\begin{figure}[t!]
        \centering
	    \includegraphics[scale=0.5]{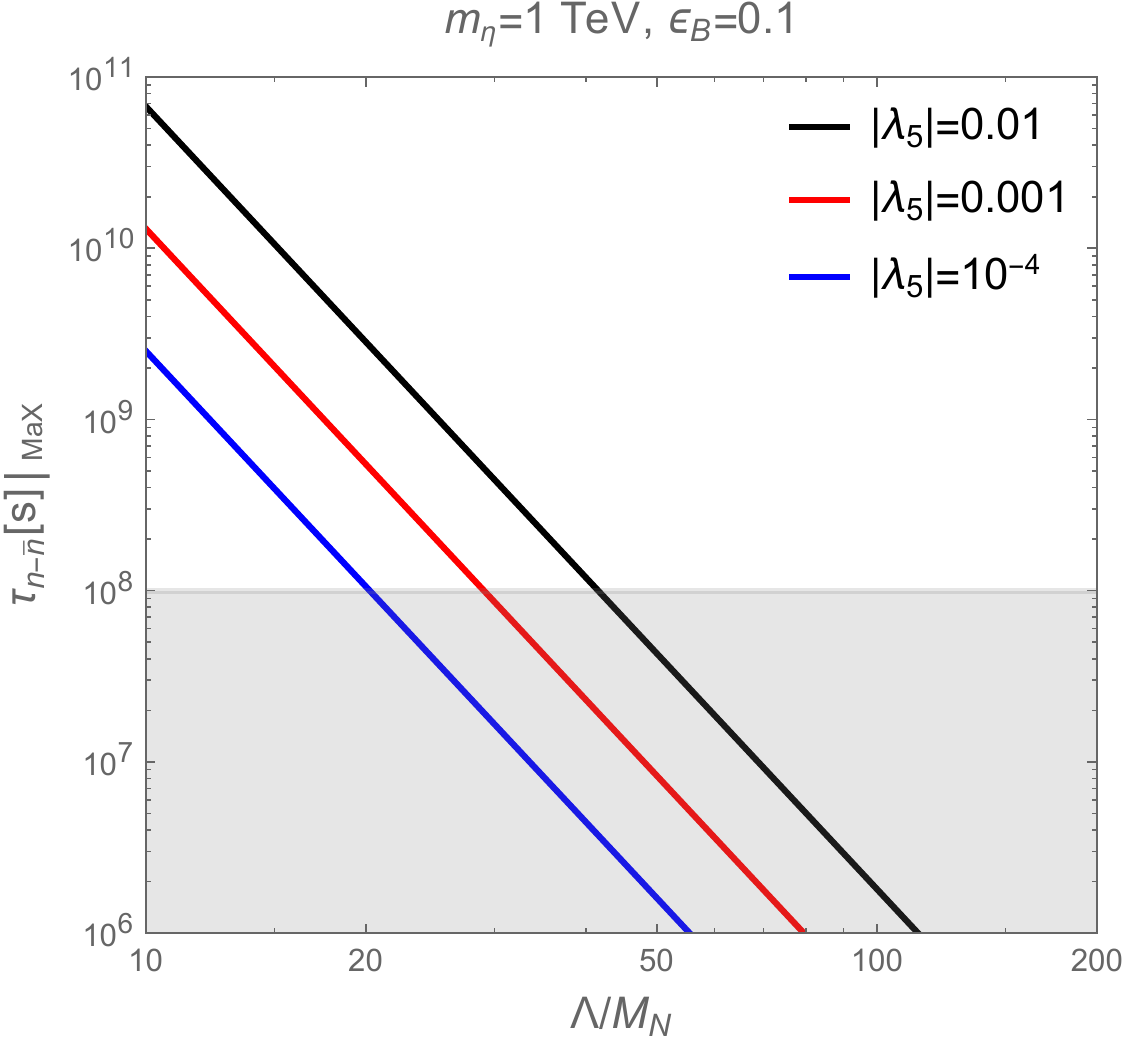}
	%\end{subfigure}
 \caption{Theoretical upper limit on $\tau_{n-\bar{n}}$ as a function of $\frac{\Lambda}{M_N}$ for $\epsilon_B=0.1$ for different values of  $| \lambda_5| =10^{-4}$.  }
\label{fig:TauUB-3}
\end{figure}

Noting the naturalness bound of $m_\eta \lesssim $1 TeV,  $\beta \equiv \frac{\Lambda}{M_N} \gg 1$, $\epsilon_B <1$, 
 and $ | \lambda_5| \ll 1$ as suggested by Fig.~\ref{fig:TauNN-1},  
we plot the theoretical upper limits on $\tau_{n-\bar{n}} $ in Figs.~\ref{fig:TauUB-1}, \ref{fig:TauUB-2} and \ref{fig:TauUB-3}. 
We find that most of the upper limits are below $10^{11}$ sec.~for a large range of parameters $\frac{\Lambda}{M_N}$, $\epsilon_B$ and  $| \lambda_5| $ in the model.  We see that the predicted upper limits on $\tau_{n-\bar{n}} $ from dilution constraint are complementary to the decay constraints. Combining the results in Fig.~\ref{fig:TauUB-2} and Fig.~\ref{fig:TauNN-1}, we see that  our model allows only a limited range of values for  $\tau_{n-\bar{n}} $, all of which is accessible to the planned ESS HIBEAM/NNBAR experiment. A negative result in this experiment therefore will rule out our model for baryogenesis. This is the main result of our paper.

%%%%%%%%%%%%%%%%%%

% \begin{eqnarray}
% \beta m_N  =\Lambda = \frac{(\alpha\lambda_5)^{1/4} m^{3/4}_Nv^{1/2}}{(748\pi^4)^{1/4} m^{1/4}_\nu}
% \end{eqnarray}
% where $\beta \gg 1$. Putting in $m_\nu\sim 5\times 10^{-11}$ GeV and $v=174$ GeV, we get
% \begin{eqnarray}
% \beta m_N = \Lambda = 300(\alpha\lambda_5)^{1/4}m^{3/4}_N,
% \end{eqnarray}
% in units of GeV. We will call $\alpha$ and $\beta$ the hierarchy factors. We will choose $\beta > 10$.
% In fig. 1 we use the upper limits on $m_N$ and $\Lambda$ for various values of the hierarchy factors to plot the maximum value of $%\tau_{n-\bar{n}}$. 
% 
% From Eq. (19), we get 
% \begin{eqnarray}
% m_N~=~\left(\frac{300}{\beta}\right)^4 \alpha |\lambda_5|
% \end{eqnarray}
% From Fig 1, this implies that $m_N \geq 56$ TeV for $\beta=10$.
%

\section{5. Avoiding baryon asymmetry wash-out} 

We have to make sure that there is no washout process that can reduce the baryon asymmetry in the model.
One process that can wash out the baryon asymmetry is $qq\to \bar{q}\bar{q}\bar{q}\bar{q}$ generated by the $\Delta B=2$ interaction, 
$ \frac{1}{\Lambda^4 M_N}u_Rd_Rd_Ru_Rd_Rd_R$. Its rate must be such that its decoupling temperature $T'_d$ is higher than the reheat temperature $T_R$. To calculate $T'_d$, we set  $ n_q(T'_d)\sigma (qq\to 4\bar{ q}) v_{rel} \simeq \frac{T^{'2}_d}{M_P}$.  
Taking $n_q(T'_d) \simeq 4\times 3\times\frac{3}{4}\frac{\zeta(3)}{ \pi^2}T^{'3}_d$ and $\sigma(qq\to 4\bar{q})\simeq PS^{(4)}\frac{T^{'8}_d}{m^2_N\Lambda^8}$, 
where we estimate the four body phase space factor $PS^{(4)}=\frac{1}{2(4\pi)^5 \Gamma(4)\Gamma(3)}\simeq 1.3\times 10^{-7}$, 
we get $T'_d$ as 
\begin{eqnarray}
  T'_d \simeq 5.8 \times \left(\frac{M_N}{M_P}\right)^{1/9} \, \beta^{8/9} \, M_N. 
\end{eqnarray}
From our results shown in Figs.~1-4, we can see $10^8 < \tau_{n-\bar{n}}[{\rm s}] \lesssim 10^{11}$ and $\beta \lesssim 50$. 
Using Eq.~(\ref{tauNN}), we find 
\begin{eqnarray}
   4.0 \times 10^5 < M_N[{\rm GeV}] \,\beta^{4/5}\lesssim 1.6 \times 10^6, 
\end{eqnarray}
which means  $M_N[{\rm GeV}]=10^4 - 10^5$ for $10< \beta \lesssim 50$. 
Hence, we find $T'_d ={\cal O}(m_N)$.  
To avoid the wash-out, the  reheat temperature $T_R$ must be less than $T'_d$. 
Our choice $T_R<T_{sph}\simeq 130$ GeV satisfies this condition.

\section{6. UV-complete model for the effective B-violating interaction} 

In this section, we present a high scale renormalizable model~\cite{Dev:2015uca} which in the low energy effective theory leads to the Lagrangian in Eq.~(\ref{eqL}).
We work within the standard model gauge group $SU(3)_c \times
SU(2)_L \times U(1)_Y$ and extend its particle content by the
addition of three RHNs ($N_i$), an extra doublet ($\eta$) and an $SU(2)_L$-singlet, color-triplet scalar ($\Delta$)
with hypercharge $Y = +2/3$ and a complex gauge singlet $\Phi$ (inflaton). We denote the quark
and lepton doublets of the SM by $Q^T_L = (u_L,d_L$) and
$L^T = (\nu_L, e_L),$ and singlets by $u_R, d_R,e_R$; the SM Higgs
doublet field is denoted by $\phi$. We impose an additional
$Z_2$-symmetry under which $Q_L, u_R, d_R,,N_a$ and $\eta$ fields are
taken to be odd, whereas $L$, $,e_R$, $\phi$ , $\Delta$ and $\Phi$ fields are taken
to be even. The gauge- and $Z_2$-invariant Yukawa interaction Lagrangian involving the leptons and the new fields $\Delta$, $\Phi$ and
$N$ in the model is given by 
\begin{eqnarray}
{\cal L}~=~ \Phi \overline{N^C} N+\overline{Q_L}\phi u_R+\overline{Q_L}\tilde\phi d_R+\bar{L}\tilde\phi e_R +\bar{L}\eta N+ h.c., 
\end{eqnarray}
where, for simplicity, we have suppressed the coupling constants as well as flavor indices.

To this if we add the coupling and mass terms of the form 
\begin{eqnarray}
{\cal L}_Y= \lambda_1 \Delta u_R d_R+\lambda_2 N\Delta^*d_R + h.c., 
\label{eq2}
\end{eqnarray}
the low energy  effective Lagrangian for $M_N, m_\eta \ll  M_{\Delta}$  has the additional terms
\begin{eqnarray}
{\cal L}^N_{eff}~=~\frac{\kappa_i}{\Lambda^2}(N_iu^cd^cd^c) +Y_D\bar{ L}\eta N+h.c.
\label{eq3}
\end{eqnarray}
where  $\Lambda^2=M^2_\Delta/(\lambda_1\lambda_2)$.
It then becomes immediately clear that combination of Eqs.~(\ref{eq2}) and (\ref{eq3}) leads to an operator of type $u_Rd_Rd_R\nu$, when $\eta$ acquires a VEV. This leads to proton decay $p\to K^+\bar\nu, \pi^+\bar{\nu}$. It is therefore essential that $\eta$ does not acquire a VEV in this model and also $m_\eta > m_p$, as is the case for our model.

%There are two models depending on how the inflaton couples to $N$ and $N'$:
%\section{7. Comments }
%
%\begin{itemize}
%
%\item Such non-thermal leptogenesis has been shown to leave its imprint on the cosmic microwave radiation~\cite{Ghoshal:2022fud}, 
%which can provide a test of this scenario.
%
%\item From the above constraints, we see that as a benchmark choice for the masses in our model, we have $m_N $ in the multi-TeV  range and $m_\eta$ in the 100 GeV range.
%
%\item For an alternative type III seesaw for neutrino masses and leptogenesis in a non-thermal set-up see~\cite{NT6}.
%
%\item For a generalized scotogenic model, where both proton decay and neutron oscillation are allowed, see \cite{Gu:2016ghu}.
%
%\end{itemize}

\section{7. Conclusion} 

In conclusion, we first show that if the heavy right handed neutrinos in the minimal scotogenic model for neutrino mass, are produced in the decay of the inflaton field and $T_R \leq 10 $ GeV, then usual leptogenesis scenario does not work to produce simultaneously the baryon asymmetry. In order to cure this problem, we introduce a dimension six effective baryon violating term involving the RHN so that its B-violating decay can produce the desired amount of baryon-anti-baryon asymmetry in the universe. An experimental signature of this model is observable baryon number violating process of neutron-anti-neutron oscillation. The model has a long lived light neutral scalar, whose decay should not affect big bang nucleosynthesis and should not dilute the produced baryon asymmetry. We find that  our model allows only a limited range of values for  $\tau_{n-\bar{n}} $, all of which is accessible to the planned ESS HIBEAM/NNBAR experiment. A negative result in this experiment will therefore rule out our model for baryogenesis. .We also present a UV-complete renormalizable high scale version of this model. 

%%%%%%%%%%%%%%%%%%%%%%%%%%%%%%%%%%%%%%%%%%%%%%%
\section*{Acknowledgement}
The work of N.O. is supported in part by the United States Department of Energy Grant 
Nos.~DE-SC0012447, DE-SC0023713, and DE-SC0026347.
%%%%%%%%%%%%%%%%%%%%%%%%%%%%%%%%%%%%%%%%%%%%%%%


\begin{thebibliography}{99}


%\cite{Fukugita:1986hr}
\bibitem{Fukugita:1986hr}
M.~Fukugita and T.~Yanagida,
%``Baryogenesis Without Grand Unification,''
Phys. Lett. B \textbf{174}, 45-47 (1986)
doi:10.1016/0370-2693(86)91126-3
%4714 citations counted in INSPIRE as of 24 Dec 2025

%\cite{Minkowski:1977sc}
\bibitem{Minkowski:1977sc}
P.~Minkowski,
%``$\mu \to e\gamma$ at a Rate of One Out of $10^{9}$ Muon Decays?,''
Phys. Lett. B \textbf{67}, 421-428 (1977)
doi:10.1016/0370-2693(77)90435-X
%5667 citations counted in INSPIRE as of 24 Dec 2025

%\cite{Mohapatra:1979ia}
\bibitem{Mohapatra:1979ia}
R.~N.~Mohapatra and G.~Senjanovic,
%``Neutrino Mass and Spontaneous Parity Nonconservation,''
Phys. Rev. Lett. \textbf{44}, 912 (1980)
doi:10.1103/PhysRevLett.44.912
%7213 citations counted in INSPIRE as of 24 Dec 2025

%\cite{GellMann:1980vs}
\bibitem{GellMann:1980vs}
M.~Gell-Mann, P.~Ramond and R.~Slansky,
%``Complex Spinors and Unified Theories,''
Conf. Proc. C \textbf{790927}, 315-321 (1979)
[arXiv:1306.4669 [hep-th]].
%4292 citations counted in INSPIRE as of 24 Dec 2025

%\cite{Yanagida:1979as}
\bibitem{Yanagida:1979as}
T.~Yanagida,
%``Horizontal gauge symmetry and masses of neutrinos,''
Conf. Proc. C \textbf{7902131}, 95-99 (1979)
KEK-79-18-95.
%2557 citations counted in INSPIRE as of 24 Dec 2025

%\cite{Glashow:1979nm}
\bibitem{Glashow:1979nm}
S.~L.~Glashow,
%``The Future of Elementary Particle Physics,''
NATO Sci. Ser. B \textbf{61}, 687 (1980)
doi:10.1007/978-1-4684-7197-7{\_}15
%913 citations counted in INSPIRE as of 24 Dec 2025

\bibitem{Buchmuller:2004nz} Buchmuller, W. and Di Bari, P. and Plumacher, M.,
    %itle = "{Leptogenesis for pedestrians}",
    %eprint = "hep-ph/0401240",
    %archivePrefix = "arXiv",
    %reportNumber = "DESY-03-100, UAB-FT-551, CERN-TH-2003-199",
    Annals Phys. 315, 305--351 (2005)  doi :10.1016/j.aop.2004.02.003.
 


%\cite{Abada:2006fw}
\bibitem{Abada:2006fw}
A.~Abada, S.~Davidson, F.~X.~Josse-Michaux, M.~Losada and A.~Riotto,
%``Flavor issues in leptogenesis,''
JCAP \textbf{04}, 004 (2006)
doi:10.1088/1475-7516/2006/04/004
[arXiv:hep-ph/0601083 [hep-ph]].
%512 citations counted in INSPIRE as of 22 Dec 2025

%\cite{Davidson:2008bu}
\bibitem{Davidson:2008bu}
S.~Davidson, E.~Nardi and Y.~Nir,
%``Leptogenesis,''
Phys. Rept. \textbf{466}, 105-177 (2008)
doi:10.1016/j.physrep.2008.06.002
[arXiv:0802.2962 [hep-ph]].
%1372 citations counted in INSPIRE as of 24 Dec 2025

%\cite{Bodeker:2020ghk}
\bibitem{Bodeker:2020ghk}
D.~Bodeker and W.~Buchmuller,
%``Baryogenesis from the weak scale to the grand unification scale,''
Rev. Mod. Phys. \textbf{93}, no.3, 3 (2021)
doi:10.1103/RevModPhys.93.035004
[arXiv:2009.07294 [hep-ph]].
%227 citations counted in INSPIRE as of 12 Dec 2025

%\cite{Lazarides:1990huy}
\bibitem{Lazarides:1990huy}
G.~Lazarides and Q.~Shafi,
%``Origin of matter in the inflationary cosmology,''
Phys. Lett. B \textbf{258}, 305-309 (1991)
doi:10.1016/0370-2693(91)91090-I
%360 citations counted in INSPIRE as of 16 Dec 2025

%\cite{Murayama:1992ua}
\bibitem{Murayama:1992ua}
H.~Murayama, H.~Suzuki, T.~Yanagida and J.~Yokoyama,
%``Chaotic inflation and baryogenesis by right-handed sneutrinos,''
Phys. Rev. Lett. \textbf{70}, 1912-1915 (1993)
doi:10.1103/PhysRevLett.70.1912
%282 citations counted in INSPIRE as of 16 Dec 2025

%\cite{Ghoshal:2022kqp}
\bibitem{Ghoshal:2022kqp}
A.~Ghoshal, R.~Samanta and G.~White,
%``Bremsstrahlung high-frequency gravitational wave signatures of high-scale nonthermal leptogenesis,''
Phys. Rev. D \textbf{108}, no.3, 035019 (2023)
doi:10.1103/PhysRevD.108.035019
[arXiv:2211.10433 [hep-ph]].
%32 citations counted in INSPIRE as of 17 Dec 2025

%\cite{Zhang:2023oyo}
\bibitem{Zhang:2023oyo}
X.~Zhang,
%``Towards a systematic study of non-thermal leptogenesis from inflaton decays,''
JHEP \textbf{05}, 147 (2024)
doi:10.1007/JHEP05(2024)147
[arXiv:2311.05824 [hep-ph]].
%12 citations counted in INSPIRE as of 16 Dec 2025

%\cite{Tao:1996vb}
\bibitem{Tao:1996vb}
Z.~j.~Tao,
%``Radiative seesaw mechanism at weak scale,''
Phys. Rev. D \textbf{54}, 5693-5697 (1996)
doi:10.1103/PhysRevD.54.5693
[arXiv:hep-ph/9603309 [hep-ph]].
%92 citations counted in INSPIRE as of 24 Dec 2025

%\cite{Ma:2006km}
\bibitem{Ma:2006km}
E.~Ma,
%``Verifiable radiative seesaw mechanism of neutrino mass and dark matter,''
Phys. Rev. D \textbf{73}, 077301 (2006)
doi:10.1103/PhysRevD.73.077301
[arXiv:hep-ph/0601225 [hep-ph]].
%1682 citations counted in INSPIRE as of 24 Dec 2025

%\cite{Barbieri:2006dq}
\bibitem{Barbieri:2006dq}
R.~Barbieri, L.~J.~Hall and V.~S.~Rychkov,
%``Improved naturalness with a heavy Higgs: An Alternative road to LHC physics,''
Phys. Rev. D \textbf{74}, 015007 (2006)
doi:10.1103/PhysRevD.74.015007
[arXiv:hep-ph/0603188 [hep-ph]].
%1073 citations counted in INSPIRE as of 19 Dec 2025

%\cite{Deshpande:1977rw}
\bibitem{Deshpande:1977rw}
N.~G.~Deshpande and E.~Ma,
%``Pattern of Symmetry Breaking with Two Higgs Doublets,''
Phys. Rev. D \textbf{18}, 2574 (1978)
doi:10.1103/PhysRevD.18.2574
%1049 citations counted in INSPIRE as of 19 Dec 2025

%\cite{Kashiwase:2012xd}
\bibitem{Kashiwase:2012xd}
S.~Kashiwase and D.~Suematsu,
%``Baryon number asymmetry and dark matter in the neutrino mass model with an inert doublet,''
Phys. Rev. D \textbf{86}, 053001 (2012)
doi:10.1103/PhysRevD.86.053001
[arXiv:1207.2594 [hep-ph]].
%82 citations counted in INSPIRE as of 10 Dec 2025

%\cite{Hugle:2018qbw}
\bibitem{Hugle:2018qbw}
T.~Hugle, M.~Platscher and K.~Schmitz,
%``Low-Scale Leptogenesis in the Scotogenic Neutrino Mass Model,''
Phys. Rev. D \textbf{98}, no.2, 023020 (2018)
doi:10.1103/PhysRevD.98.023020
[arXiv:1804.09660 [hep-ph]].
%102 citations counted in INSPIRE as of 24 Dec 2025

%\cite{Borah:2018rca}
\bibitem{Borah:2018rca}
D.~Borah, P.~S.~B.~Dev and A.~Kumar,
%``TeV scale leptogenesis, inflaton dark matter and neutrino mass in a scotogenic model,''
Phys. Rev. D \textbf{99}, no.5, 055012 (2019)
doi:10.1103/PhysRevD.99.055012
[arXiv:1810.03645 [hep-ph]].
%92 citations counted in INSPIRE as of 22 Dec 2025

%\cite{Sarma:2020msa}
\bibitem{Sarma:2020msa}
L.~Sarma, P.~Das and M.~K.~Das,
%``Scalar dark matter and leptogenesis in the minimal scotogenic model,''
Nucl. Phys. B \textbf{963}, 115300 (2021)
doi:10.1016/j.nuclphysb.2020.115300
[arXiv:2004.13762 [hep-ph]].
%22 citations counted in INSPIRE as of 22 Dec 2025

%\cite{Racker:2024fpn}
\bibitem{Racker:2024fpn}
J.~Racker,
%``Low-scale leptogenesis in the scotogenic model: Spectator processes and benchmark points,''
Phys. Rev. D \textbf{111}, no.8, L081301 (2025)
doi:10.1103/PhysRevD.111.L081301
[arXiv:2411.15120 [hep-ph]].
%3 citations counted in INSPIRE as of 17 Dec 2025

%\cite{Avila:2025qsc}
\bibitem{Avila:2025qsc}
I.~M.~{\'A}vila, A.~Karan, S.~Mandal, S.~Sadhukhan and J.~W.~F.~Valle,
%``Dark matter as the source of neutrino mass: theory overview and experimental prospects,''
[arXiv:2506.24027 [hep-ph]].
%4 citations counted in INSPIRE as of 24 Dec 2025

%\cite{Dev:2015uca}
\bibitem{Dev:2015uca}
P.~S.~B.~Dev and R.~N.~Mohapatra,
%``TeV scale model for baryon and lepton number violation and resonant baryogenesis,''
Phys. Rev. D \textbf{92}, no.1, 016007 (2015)
doi:10.1103/PhysRevD.92.016007
[arXiv:1504.07196 [hep-ph]].
%67 citations counted in INSPIRE as of 10 Dec 2025

%\cite{Davoudiasl:2015jja}
\bibitem{Davoudiasl:2015jja}
H.~Davoudiasl and Y.~Zhang,
%``Baryon Number Violation via Majorana Neutrinos in the Early Universe, at the LHC, and Deep Underground,''
Phys. Rev. D \textbf{92}, no.1, 016005 (2015)
doi:10.1103/PhysRevD.92.016005
[arXiv:1504.07244 [hep-ph]].
%24 citations counted in INSPIRE as of 10 Dec 2025

\bibitem{Kuzmin:1970nx} V. Kuzmin,   Pisma Zh.Eksp.Teor.Fiz. 12 (1970) 6, 335-337, JETP Lett. 12 (1970) 6, 228.

%\bibitem{Glashow:1979nm} S. L. Glashow, NATO Sci.Ser.B 61 (1980) 687 • Contribution to: Cargese Summer Institute: Quarks and Leptons.

\bibitem{Mohapatra:1980qe} Rabindra N. Mohapatra and R.E. Marshak, Phys. Rev. Lett. \textbf{44}, (1980) 1316-1319, Phys.Rev.Lett. 44 (1980) 1643 (erratum).  doi :10.1103/PhysRevLett.44.1316.

%\cite{Babu:2006xc}
\bibitem{Babu:2006xc}
K.~S.~Babu, R.~N.~Mohapatra and S.~Nasri,
%``Post-Sphaleron Baryogenesis,''
Phys. Rev. Lett. \textbf{97}, 131301 (2006)
doi:10.1103/PhysRevLett.97.131301
[arXiv:hep-ph/0606144 [hep-ph]].
%121 citations counted in INSPIRE as of 10 Dec 2025

%\cite{Babu:2013yca}
\bibitem{Babu:2013yca}
K.~S.~Babu, P.~S.~Bhupal Dev, E.~C.~F.~S.~Fortes and R.~N.~Mohapatra,
%``Post-Sphaleron Baryogenesis and an Upper Limit on the Neutron-Antineutron Oscillation Time,''
Phys. Rev. D \textbf{87}, no.11, 115019 (2013)
doi:10.1103/PhysRevD.87.115019
[arXiv:1303.6918 [hep-ph]].
%99 citations counted in INSPIRE as of 04 Dec 2025


%\cite{Okada:2010jf}
\bibitem{Okada:2010jf}
N.~Okada, M.~U.~Rehman and Q.~Shafi,
%``Tensor to Scalar Ratio in Non-Minimal $\phi^4$ Inflation,''
Phys. Rev. D \textbf{82}, 043502 (2010)
doi:10.1103/PhysRevD.82.043502
[arXiv:1005.5161 [hep-ph]].
%139 citations counted in INSPIRE as of 04 Dec 2025

%\cite{Bezrukov:2007ep}
\bibitem{Bezrukov:2007ep}
F.~L.~Bezrukov and M.~Shaposhnikov,
%``The Standard Model Higgs boson as the inflaton,''
Phys. Lett. B \textbf{659}, 703-706 (2008)
doi:10.1016/j.physletb.2007.11.072
[arXiv:0710.3755 [hep-th]].
%2209 citations counted in INSPIRE as of 25 Dec 2025

%\cite{Planck:2018vyg}
\bibitem{Planck:2018vyg}
N.~Aghanim \textit{et al.} [Planck],
%``Planck 2018 results. VI. Cosmological parameters,''
Astron. Astrophys. \textbf{641}, A6 (2020)
[erratum: Astron. Astrophys. \textbf{652}, C4 (2021)]
doi:10.1051/0004-6361/201833910
[arXiv:1807.06209 [astro-ph.CO]].
%20060 citations counted in INSPIRE as of 25 Dec 2025

%\cite{Lundstrom:2008ai}
\bibitem{Lundstrom:2008ai}
E.~Lundstrom, M.~Gustafsson and J.~Edsjo,
%``The Inert Doublet Model and LEP II Limits,''
Phys. Rev. D \textbf{79}, 035013 (2009)
doi:10.1103/PhysRevD.79.035013
[arXiv:0810.3924 [hep-ph]].
%314 citations counted in INSPIRE as of 19 Dec 2025

%\cite{Ilnicka:2015jba}
\bibitem{Ilnicka:2015jba}
A.~Ilnicka, M.~Krawczyk and T.~Robens,
%``Inert Doublet Model in light of LHC Run I and astrophysical data,''
Phys. Rev. D \textbf{93}, no.5, 055026 (2016)
doi:10.1103/PhysRevD.93.055026
[arXiv:1508.01671 [hep-ph]].
%181 citations counted in INSPIRE as of 08 Dec 2025

%\cite{Lahiri:2025opz}
\bibitem{Lahiri:2025opz}
J.~Lahiri, T.~Robens and K.~Rolbiecki,
%``Constraining the Inert Doublet Model at the LHC,''
[arXiv:2511.23133 [hep-ph]].
%0 citations counted in INSPIRE as of 25 Dec 2025

%\cite{Rinaldi:2019thf}
\bibitem{Rinaldi:2019thf}
E.~Rinaldi, S.~Syritsyn, M.~L.~Wagman, M.~I.~Buchoff, C.~Schroeder and J.~Wasem,
%``Lattice QCD determination of neutron-antineutron matrix elements with physical quark masses,''
Phys. Rev. D \textbf{99}, no.7, 074510 (2019)
doi:10.1103/PhysRevD.99.074510
[arXiv:1901.07519 [hep-lat]].
%70 citations counted in INSPIRE as of 10 Dec 2025

%\cite{Rao:1982gt}
\bibitem{Rao:1982gt}
S.~Rao and R.~Shrock,
%``$n \leftrightarrow \bar{n}$ Transition Operators and Their Matrix Elements in the {MIT} Bag Model,''
Phys. Lett. B \textbf{116}, 238-242 (1982)
doi:10.1016/0370-2693(82)90333-1
%137 citations counted in INSPIRE as of 04 Dec 2025

%\cite{Santoro:2023izd}
\bibitem{Santoro:2023izd}
V.~Santoro, D.~Milstead, P.~Fierlinger, W.~M.~Snow, J.~Amaral, J.~Barrow, M.~Bartis, P.~Bentley, L.~Bj{\"o}rk and G.~Brooijmans, \textit{et al.}
%``The HIBEAM instrument at the European spallation source,''
J. Phys. G \textbf{52}, no.4, 040501 (2025)
doi:10.1088/1361-6471/adc8c2
[arXiv:2311.08326 [physics.ins-det]].
%12 citations counted in INSPIRE as of 22 Dec 2025

%\cite{Pierce:2007ut}
\bibitem{Pierce:2007ut}
A.~Pierce and J.~Thaler,
%``Natural Dark Matter from an Unnatural Higgs Boson and New Colored Particles at the TeV Scale,''
JHEP \textbf{08}, 026 (2007)
doi:10.1088/1126-6708/2007/08/026
[arXiv:hep-ph/0703056 [hep-ph]].
%201 citations counted in INSPIRE as of 19 Dec 2025

%\cite{ATLAS:2024lda}
\bibitem{ATLAS:2024lda}
G.~Aad \textit{et al.} [ATLAS],
%``The quest to discover supersymmetry at the ATLAS experiment,''
Phys. Rept. \textbf{1116}, 261-300 (2025)
doi:10.1016/j.physrep.2024.09.010
[arXiv:2403.02455 [hep-ex]].
%65 citations counted in INSPIRE as of 24 Dec 2025

%\cite{LZ:2024zvo}
\bibitem{LZ:2024zvo}
J.~Aalbers \textit{et al.} [LZ],
%``Dark Matter Search Results from 4.2{\,}{\,}Tonne-Years of Exposure of the LUX-ZEPLIN (LZ) Experiment,''
Phys. Rev. Lett. \textbf{135}, no.1, 011802 (2025)
doi:10.1103/4dyc-z8zf
[arXiv:2410.17036 [hep-ex]].
%292 citations counted in INSPIRE as of 25 Dec 2025

%\cite{Chowdhury:2015yja}
\bibitem{Chowdhury:2015yja}
D.~Chowdhury and O.~Eberhardt,
%``Global fits of the two-loop renormalized Two-Higgs-Doublet model with soft Z$_{2}$ breaking,''
JHEP \textbf{11}, 052 (2015)
doi:10.1007/JHEP11(2015)052
[arXiv:1503.08216 [hep-ph]].
%109 citations counted in INSPIRE as of 17 Dec 2025

\bibitem{Baldo-Ceolin:1994hzw} M. Baldo-Ceolin et al., Z. Phys. C \textbf{63}, 409 (1994) 
doi :10.1007/BF01580321

\bibitem{Super-Kamiokande:2020bov} K. Abe et al.  Phys.Rev.D 103 (2021) 1, 012008 • e-Print: 2012.02607 [hep-ex]

%\cite{Ghoshal:2022fud}
%\bibitem{Ghoshal:2022fud}
%A.~Ghoshal, D.~Nanda and A.~K.~Saha,
%``CMB imprints of high scale non-thermal leptogenesis,''
%Phys. Lett. B \textbf{849}, 138484 (2024)
%doi:10.1016/j.physletb.2024.138484
%[arXiv:2210.14176 [hep-ph]].
%19 citations counted in INSPIRE as of 17 Dec 2025

%\cite{Gu:2016ghu}
%\bibitem{Gu:2016ghu}
%P.~H.~Gu, E.~Ma and U.~Sarkar,
%``Connecting Radiative Neutrino Mass, Neutron-Antineutron Oscillation, Proton Decay, and Leptogenesis through Dark Matter,''
%Phys. Rev. D \textbf{94}, no.11, 111701 (2016)
%doi:10.1103/PhysRevD.94.111701
%[arXiv:1608.02118 [hep-ph]].
%30 citations counted in INSPIRE as of 04 Dec 2025

\end{thebibliography}
\end{document}